\documentclass[aps,prb]{revtex4}

\usepackage{graphicx,amsmath,amssymb,bm}

\begin{document}

\title{Microscopic free energy functional of superconductive amplitude and phase: Superfluid density in disordered superconductors}
\author{Sudhansu S. Mandal$^{1,2}$ and T. V. Ramakrishnan$^3$ }
\affiliation{$^1$Department of Physics, Indian Institute of Technology, Kharagpur 721302, India\\ 
	$^2$Centre for Theoretical Studies, Indian Institute of Technology, Kharagpur 721302, India	\\
	$^3$Department of Physics, Indian Institute of Science, Bangalore 5600012, India}

\date{\today}

\begin{abstract}
	Recent experiments on disordered superconductors find that the superfluid density $n_s(T)$ decreases dramatically and characteristically with disorder, differently from what is expected for a mean order parameter field or BCS, amplitude only, picture for the superconductor. We describe here a new microscopic free energy functional which explicitly describes its dependence on both the amplitude and phase of superconducting order, in a gauge invariant manner.  We use this here in an approximation of noninteracting phase fluctuations (Gaussian or harmonic approximation) to obtain $n_s(T)$ in the presence of (static) disorder. We compare our results successfully with experiment.

\end{abstract}

\maketitle

\section{Introduction}

Superconductivity is identified with a nonzero complex order parameter 
 $\Psi (x) =  \Delta (x) \exp[i\phi(x)]$ where $(x)$ is shorthand for spatial coordinate ${\rm\bf r}$ and imaginary time $\bar{\tau}$,  $ \Delta (x)$ is the amplitude of the order parameter and $\phi (x)$ is its phase. The microscopic theory of superconductivity \cite{BCS,Gorkov} identifies it with the statistical average of the electron Cooper pair density, namely $\Psi(x)=<\psi_\uparrow^\dagger (x)\psi_\downarrow^\dagger (x)>$, $\psi_\uparrow^\dagger (x)$ being the up spin electron creation operator. If the system has static nonmagnetic disorder which preserves time reversal invariance, the condition for Cooper pairing of electrons in time reversed states does not change provided the effective pairing attraction does not. This is Anderson's theorem \cite{Anderson}, and  consequently, in the Bardeen-Cooper-Schrieffer (BCS) mean field theory the superconducting transition temperature $T_c$ does not change \cite{footnote_anderson}, as broadly confirmed by experiment. Recent work on disordered superconductors shows however that both $T_c$ and superfluid density $n_s (T)$ decrease with disorder in a characteristic way \cite{Pratap1,Pratap2}, different from what is expected in the BCS theory. The single particle density of states exhibits a pseudogap above $T_c$, somewhat like what is seen in cuprate superconductors \cite{Review1,Review2,Review3}. Moreover, there seems to be a new state of quantum matter, the failed superconductor state which occurs for relatively weak disorder or no disorder \cite{Kivelson}. This state is most likely characterized by Cooper pairs whose phases are not mutually coherent. In all these systems, disorder is not large enough for effects of Anderson localization to be significant so that issues such as the superconductor-insulator transition due to disorder or the `superinsulator' phase are not relevant. It seems quite likely that all these phenomena are connected with the effect of the phase $\phi (x)$, affected by disorder or interactions. Therefore, there is need for a description of a superconductor which involves the phase $\phi(x)$ explicitly, in addition to the amplitude $\Delta(x)$ and for the use of such a description to calculate physical properties.

 The celebrated Ginzburg Landau \cite{GL} functional and its microscopic derivation by Gor’kov \cite{Gorkov} describe power series expansions of the free energy of a superconductor as a functional of  $\Psi (x)$. Such an approach is therefore restricted to temperatures in the vicinity of $T_c$ where $\Psi (x)$ is small. Other theories developed so far are based on either microscopic amplitude-only \cite{A-G,Kogan} or phase-only \cite{TVR,Benfatto1} approaches, or phenomenological approaches (see, e.g, Refs.\onlinecite{Benfatto2} and \onlinecite{Mirlin}).

 In this paper, for the first time, we obtain microscopically the free-energy for a superconductor as 
  a  functional of both the  amplitude and phase of the superconducting order parameter at all temperatures. We outline here a simple general approach which directly expresses the free energy of the superconductor in terms of the electron pair phase $\phi(x)$ in addition to the amplitude $\Delta(x)$. The approach can be motivated starting from an attractive pairing interaction between electrons and doing an exact Hubbard Stratonovich transformation on it to express the free energy as that of electrons moving in a time and space dependent, complex pairing potential which can be identified with the order parameter. Using this, the phase-dependent Hamiltonian is obtained as a function of the gauge invariant superfluid velocity ${\rm \bf v}_s (x) = (1/m)(\hbar\bm{\nabla}\phi - (2e/c)\bm{A})$. At this stage, the Hamiltonian depends on the electron degrees of freedom in addition to superconducting amplitude and phase degrees of freedom. On integrating out the electron degrees of freedom, we have a functional explicitly and separately of the order parameter amplitude, as well as of its phase (actually of the superfluid velocity which, as mentioned above, is linear to the gradient of the phase).
 
We apply this approach here in the simplest Gaussian or harmonic approximation for phase fluctuations 
 (in which they are noninteracting) to determine the superfluid density $n_{s}(T)$ at all temperatures. In this approximation, the free energy is quadratic in phase fluctuations and the coefficient depends \textit{only} on the amplitude of the order parameter. 
 The coefficient of the contribution  quadratic in ${\rm \bf v}_s$, in the uniform and static limit is by definition  $(1/2)\rho_{s}$, where $\rho_s$ is the superfluid stiffness (Section II and see also Ref.\onlinecite{TVR}). This is proportional to the superfluid density, $n_s$, via the relation $n_s = 4\rho_s/m$. 
 
There is a microscopic theory of the linear electromagnetic response of superconducting alloys \cite{A-G} 
 for determining the superfluid density; here the alloy is treated in the BCS mean field approximation with a uniform amplitude $\Delta$.  This is  restricted to only  $T=0$ because of its complex approach. There are phenomenological extensions to $T \neq 0$ (a relatively recent example is \cite{Kogan}. Our approach here enables us to determine $n_s(T)$, microscopically, for all $T$.
    
Using the above approximation to the new functional, we obtain in the subsequent section 
 (Section III) results for $n_s(T=0)$ 
 and $n_s (T/T_{c})$ as a function of static nonmagnetic disorder (characterized by a relaxation time $\tau$). We apply methods\cite{Ambegaokar} standard for a many electron system with random impurities leading to a nonzero $\tau^{-1}$, and with a zero range BCS pairing attraction. The disorder dependence of $n_s(T=0)$ is explicitly calculated in the relaxation time approximation; the ground state or $T=0$ London value $n$ for $n_{s}$, is recovered in the clean limit $(\tau^{-1} \to 0)$. We also obtain the dependence of $n_{s}(T)$ on $(T/T_c)$ and show that the other limiting behavior near $T_c$ (namely its going to zero) is also correctly obtained. In section III, we also compare our results broadly with experiment \cite{Pratap1} on disordered superconductors, and obtain agreement both as to the trend of $n_s(T=0)$ as a function of disorder (it decreases!) and the size. We show analytically that it decreases linearly with disorder for small disorder as is seen in experiment, and as predicted in Ref.\onlinecite{A-G}. We obtain closed form expressions for all $T$, and for $\Delta_0\tau < 1$ as well as for $\Delta_0\tau > 1$ (but $\epsilon_{F}\tau << 1$), where $\Delta_0$ is the BCS gap at $T=0$ and $\epsilon_F$ is the Fermi energy. 
 
 A discussion of the novelty of the general microscopic approach in obtaining free energy functional of amplitude and phase is given in section IV. 
  Some possible future directions including application to the failed superconductor situation (where $n_s(T=0)$ vanishes at $T=0$) are also suggested there.

\section{General Formalism}  
 
We consider electrons in a nonmagnetic random potential $V({\rm \bf r})$ interacting via a zero range BCS attractive potential of strength $g$. The system Hamilitonian is given by
  \begin{eqnarray}
	H  &=& \int d{\rm \bf r} \left[ \sum_{\sigma}  \psi_\sigma^\dagger ({\rm \bf r}) \left( \frac{1}{2m} (-i\hbar \bm{\nabla} - \frac{e}{c} \bm{A})^2 + V({\bm r}) -\mu \right) \psi_\sigma ({\bm r}) \right. \nonumber \\
 && \left. -g \psi_\uparrow^\dagger ({\rm \bf r})\psi_\downarrow^\dagger ({\rm \bf r}) \psi_\downarrow ({\rm \bf r}) \psi_\uparrow ({\rm \bf r}) \right] 
	\label{Hamiltonian}
\end{eqnarray}
where $\psi_\sigma^\dagger ({\rm \bf r})$ and $\psi_\sigma ({\rm \bf r})$ respectively represent fermionic creation and destruction operators with spin $\sigma = \uparrow $ or $\downarrow$, a chemical potential $\mu$ for the fixed density of electrons, and $\bm{A}$ is the vector potential. The corresponding partition function in terms of coherent state path integrals of the Grassmannian fields $\psi_\sigma (x)$ and $\bar{\psi}_\sigma (x)$ is given by
\begin{equation}
{\cal Z} = \int {\cal D}\{\psi_\sigma, \bar{\psi}_\sigma\}\, \exp[- S_E]
\end{equation}
where $x\equiv ({\rm \bf r},\bar{\tau})$ represents both the coordinate vector ${\rm \bf r}$ and imaginary time $\bar{\tau}$ and the Euclidean action ${\cal S}_E$ reads
\begin{eqnarray}
S_E &=& \int_0^\beta d\bar{\tau} \int d{\rm \bf r} \left[  \sum_{\sigma}  \bar{\psi}_\sigma (x)\left( \hbar\partial_{\bar{\tau}}   +\frac{1}{2m}(-i\hbar\bm{\nabla} -e \bm{A})^2 + V({\bm r})-\mu \right)    \psi_\sigma (x) \right. \nonumber \\
&&\left.  -g \bar{\psi}_\uparrow (x)\bar{\psi}_\downarrow (x) \psi_\downarrow (x) \psi_\uparrow (x) \right]
\label{Action_1}
\end{eqnarray}
with inverse temperature $\beta = 1/k_{_B}T$.

 By introducing bosonic fields $\Psi (x)$ and $\Psi^\ast(x)$ via the Hubbard-Stratonovich transformation for decomposing the quartic term in the Grassmannian fields in Eq.(\ref{Action_1}), one finds the partition function
\begin{equation}
	{\cal Z} = \int {\cal D}\{ \psi_\sigma, \bar{\psi}_\sigma,\Psi,\Psi^\ast\} \exp[-{\cal S}_E]
\end{equation}
in terms of path integrals over Grassmanian variables $\psi_\sigma$ and $\bar{\psi}_\sigma$ and bosonic variables $\Psi$ and $\Psi^\ast$; the renormalized Euclidean action ${\cal S}_E$ is then given by
\begin{eqnarray}
	{\cal S}_E &=& \int_0^\beta d\bar{\tau} \int d{\rm \bf r} \left[  \sum_{\sigma}  \bar{\psi}_\sigma (x)\left( \hbar\partial_{\bar{\tau}}   +\frac{1}{2m}(-i\hbar \bm{\nabla}-e\bm{A})^2 + V({\bm r})-\mu \right)  \psi_\sigma (x)	 \right. \nonumber \\
	&+& \left.  \Psi^\ast (x)  \psi_\downarrow (x) \psi_\uparrow (x) + \Psi (x)  \bar{\psi}_\uparrow (x) \bar{\psi}_\downarrow (x) 
	+  \frac{\vert \Psi (x) \vert^2}{g} \right]
	\label{Action_2}
\end{eqnarray}
where the complex $\Psi (x) = \Delta (x) e^{i\phi(x)}$ is identified with the superconducting order parameter or pair-potential of amplitude $\Delta (x)$ having Gaussian probability distribution $\exp[-\Delta^2 (x)/g]$ and phase $\phi (x)$. One can redefine $\psi_\sigma (x)$ for making the pair-potential real {\it a la} the BCS mean field pair-potential. However, this is not unique, and the manner in which $\phi(x)$ can be connected in different ways with the members of the electron pair have been discussed by several authors \cite{GL,S-G,Anderson2,Franz}. Here we use a symmetric gauge \cite{footnote}, namely transform the Grassmannian fields as $\psi_\sigma (x) = \tilde{\psi}_\sigma (x) \exp[i\phi(x)/2]$. We thus find ${\cal S}_E \to {\cal S}_{\rm eff} = {\cal S}_0 + {\cal S}_\phi$ with  the amplitude-only action
\begin{eqnarray}
{\cal S}_0 &=&  \int_0^\beta \int d{\rm \bf r} \left[ \sum_{\sigma}\bar{\tilde{\psi}}_\sigma (x) \left( \hbar \frac{\partial}{\partial\bar{\tau}} +\frac{{\rm \bf p}^2}{2m} + V({\rm \bf r}) -\mu\right) \tilde{\psi}_\sigma (x)  \right. \nonumber \\
& & \left. + \Delta(x) \left(  \tilde{\psi}_\downarrow (x) \tilde{\psi}_\uparrow (x) +  \bar{\tilde{\psi}}_\uparrow (x) \bar{\tilde{\psi}}_\downarrow (x) \right)
+  \frac{\Delta^2 (x)}{g} \right]
\label{Action_0}
\end{eqnarray}
and phase-dependent action
\begin{equation}
{\cal S}_\phi = \frac{1}{4} \int_0^\beta d\bar{\tau}\int d{\rm \bf r} \sum_\sigma \bar{\tilde{\psi}}_\sigma(x) \left[ 2i\hbar  \left(\frac{\partial \phi}{\partial \bar{\tau}}\right) + \left({\rm \bf p}\bm{\cdot} {\rm \bf v}_s +{\rm \bf v}_s \bm{\cdot} {\rm \bf p}\right) +\frac{m}{2} {\rm \bf v}_s^2 \right] \tilde{\psi}_\sigma (x) \,.
\label{Action_phi}
\end{equation}
where momentum operator ${\rm \bf p}=-i\hbar \bm{\nabla}$ and superfluid velocity ${\rm \bf v}_s = (1/m) \left[ \hbar \bm{\nabla}\phi- 2(e/c) \bm{A} \right]$.

 We now integrate over Grassmannian fields $\tilde{\psi}_\sigma$ and $\bar{\tilde{\psi}}_\sigma$ to determine the partition function ${\cal Z} = \exp[ -\beta ({\cal F}_0 + {\cal F}_\phi)]$. The amplitude-only free energy is given by
\begin{equation}
{\cal F}_0 (\Delta (x)) = \frac{1}{\beta} \left[ \int_0^\beta d\bar{\tau}\int d{\rm \bf r} \frac{\Delta^2(x)}{g} - {\rm Tr}\, \ln \left( \hbar\frac{\partial}{\partial \bar{\tau}} + {\cal H}_0  \right)\right]
\label{F0_1}
\end{equation}
where the Hamiltonian 
\begin{equation}
{\cal H}_0 = \sigma_3 \left( \frac{1}{2m} {\rm \bf p}^2 + V({\rm \bf r}) -\mu \right)  -\sigma_1 \Delta (x)
\label{H0_1} 
\end{equation}
in the Nambu spinor basis $(\bar{\tilde{\psi}}_\uparrow ,\, \tilde{\psi}_\downarrow )$ represented by Pauli matrices. Here ${\rm Tr}$ represents trace over space-time as well as over spin matrices. The Hamiltonian (\ref{H0_1}) describes the BCS Hamiltonian with random potential when $\Delta (x)$ becomes space-time independent. The phase-dependent free energy is found to be
\begin{equation}
{\cal F}_\phi = -\frac{1}{\beta}  \sum_{n=1}^\infty (-1)^{n+1}\frac{1}{n}   {\rm Tr}\left[ \left( \left( \hbar\frac{\partial}{\partial \bar{\tau}} + {\cal H}_0  \right)^{-1} {\cal H}_\phi \right)^n \right]
\label{F_phi_1}
\end{equation}
where the phase Hamiltonian
\begin{equation}
{\cal H}_\phi = \frac{1}{4} \left[ \left( 2i\hbar  \left(\frac{\partial \phi}{\partial \bar{\tau}}\right) +\frac{m}{2} {\rm \bf v}_s^2 \right) \sigma_3  +\left({\rm \bf p}\bm{\cdot} {\rm \bf v}_s +{\rm \bf v}_s \bm{\cdot} {\rm \bf p}\right)\right]
\label{H_phi_1}
\end{equation}
We note that ${\cal H}_\phi$ is an emergent Hamiltonian involving explicitly the fluctuating phase of $\Psi (x)$. 
 This will have nontrivial consequences. We show below, as an example, that this provides a direct way of determining superfluid stiffness. 
  
At this stage, namely at the stage of the Hamiltonians and free energy functionals, namely of 
 equations (\ref{H0_1}) with (\ref{F0_1}) and (\ref{H_phi_1}) with (\ref{F_phi_1}) respectively, the order $\Psi$ and its phase $\phi$ can have arbitrary space and time dependences; the functional is completely general, while at the same time depending separately on the phase as well as the amplitude. Thus it has the wide applicability of the original Ginzburg Landau functional; it can also be used when the order is strongly inhomogeneous in space and time.

In this paper, we calculate the superfluid stiffness of the superconductor in thermal equilibrium. 
 This is a dc property, proportional to the coefficient of ${\rm \bf v}_s^2$ for the equiliibrium superconductor in the momentum ${\rm \bf q}\rightarrow 0$ and frequency $\omega\rightarrow 0$ or uniform, static, limit. We calculate this dc property here in the harmonic approximation, where fluctuations with different ${\rm \bf q}$ and $\omega$ values are independent, so that space or time dependence in ${\rm \bf v}_s$ is irrelevant. Since the momentum-momentum correlation function in the superconductor, which is the coefficient of ${\rm \bf v}_s^2$,  involves the order parameter amplitude $\Delta(x)$, spatial and temporal fluctuations in it could affect our estimate. We ignore these, and our calculations of the stiffness assume a static, uniform $\Delta$ (this is the BCS approximation). The general reason for ignoring spatial fluctuations is that these have a length scale of order of the coherence length $\xi$ which is always much larger than the inverse of the Fermi length, namely $k_{_F}^{-1}$ so that the spatial order parameter fluctuations involve a small parameter $(k_{_F}\xi)^{-1}<<1$. The temporal fluctuations have a natural time scale $|\Delta|^{-1}$, which is much larger than the characteristic electronic time scale $\epsilon_{F}^{-1}$. We therefore also ignore the time dependence of $\Delta(x)$. However, the formalism is general enough to allow one to calculate the effect of inhomogeneities in the order parameter connected with static randomness, namely the term in the free energy which arises as a result of the cross correlation between the potential fluctuation and the spatial dependence of the superconducting order. (This is may be the origin of static patches \cite{Ghosal} where the superconducting amplitude is enhanced because the random potential favors them energetically). We can also calculate the effect of Gaussian level spatial fluctuations in $\Delta({\rm \bf r})$.

\subsection{Disorder Averaged Green's Function and ${\cal F}_0$ }

The matrix differential operator, $[\hbar (\partial /\partial \bar{\tau}) +{\cal H}_0]$, satisfies 
 the equation of motion of the Green's function in the Nambu-spinor basis as
 \begin{equation}
-\left( \hbar\frac{\partial}{\partial \bar{\tau}} + {\cal H}_0  \right) {\cal G}(x,x') =\delta (x-x')
 \end{equation}
 where
 \begin{equation}
 {\cal G}(x,x') = \left[ \begin{array}{ll} G(x,x') & F(x,x') \\ F^\dagger (x,x') & -G(x',x) \end{array} \right] 
 \end{equation}
 with $G(x,x')= - \langle T_{\bar{\tau}} \tilde{\psi}_\uparrow (x) \bar{\tilde{\psi}}_\uparrow (x') \rangle_0$ and $ F(x,x')= - \langle T_{\bar{\tau}} \tilde{\psi}_\uparrow (x)\tilde{\psi}_\downarrow (x')\rangle_0$
 being the normal and anomalous Green's functions respectively where $T_{\bar{\tau}}$ represents time-ordering and $\langle \cdots \rangle_0$ represents statistical average with respect to ${\cal F}_0$.

In the absence of disorder potential, the BCS limit  corresponds to the assumption 
 $\Delta(x) = \Delta$,   i.e., space-time independent. The disorder configuration averaged Green's function $\langle {\cal G}
	({\rm \bf r}\bar{\tau}, {\rm \bf r}'\bar{\tau}') \rangle_{\rm dis} \equiv{\cal G}({\rm \bf r}-{\rm \bf r}',\bar{\tau}-\bar{\tau}') $ in the Fourier basis for frequency and momentum is thus obtained  as 
	\begin{equation}
	{\cal G} (i\omega_n, {\rm \bf k}) = \frac{i\tilde{\omega}_n \sigma_0 +\xi_{{\rm \bf k}} \sigma_3 + \sigma_1   \tilde{ \Delta}  }{(i\tilde{\omega}_n)^2 -\xi_{{\rm \bf k}}^2 -  \tilde{\Delta}^2} .
	\label{Green}
	\end{equation}
	The fermionic Matsubara frequency $\omega_n = (2n+1)\pi T$ and the BCS pair-amplitude $\Delta$ are related \cite{A-G,Ambegaokar} with their normalized counterparts as  
	\begin{equation}
	\frac{\tilde{\omega}_n}{\omega_n} = \frac{ \tilde{\Delta}}{\Delta } = 1+ \frac{1}{2\tau \sqrt{ \Delta^2 +\omega_n^2}} \, .
	\label{frequency_relation}
	\end{equation}
	This relation is determined by evaluating the self-energy of the quasiparticles, namely, 
	\begin{equation}
	\Sigma (i\omega_n,{\rm \bf k}) = \frac{1}{2\pi\nu \tau} \int \frac{d^3{\rm \bf q}}{(2\pi)^3} \sigma_3 {\cal G}(i\omega_n,{\rm \bf k-q})\sigma_3
	\end{equation}
  where the factor $1/(2\pi\nu\tau)$ is realized from the white noise disorder potential given by $\langle V({\rm \bf r}) V({\rm \bf r}') \rangle = \frac{1}{2\pi \nu \tau} \delta ({\rm \bf r}-{\rm \bf r}')$ with 
	$\nu$ being the density of states of electrons of each spin at the Fermi energy and $\tau$ the momentum relaxation time for elastic scattering.

 Upon averaging over disorder configurations, Eq.(\ref{F0_1}) reduces to
 \begin{eqnarray}
 {\cal F}_0 (\Delta) &=& \int d{\rm \bf r} \left[ \frac{\Delta^2}{g} - {\rm tr} \,\ln\left( \frac{1}{\beta}\sum_{\omega_n} \int \frac{d{\rm \bf k}}{(2\pi)^3} {\cal G}^{-1} (i\omega_n,\, {\rm \bf k}) \right) \right] \\
 &=& \int  d{\rm \bf r} \left[ \frac{\Delta^2}{g} - 4\pi \nu T \sum_{\omega_n>0} \left( \sqrt{\omega_n^2 +  \Delta^2} -  \omega_n \right)    \right]
 \end{eqnarray}
 where ${\rm tr}$ represents trace only in the Nambu basis,
 and $\omega_n$ has been subtracted in the equation above to keep the terms which survive only when $\Delta \neq 0$.
 In other words, we ignore part of the free energy in the absence of $\Delta$, i.e., normal state free energy.  
 This is the BCS mean field free energy which does not get renormalized due to disorder, as is well known in accordance with Anderson's theorem \cite{Anderson}.

 The approximation $\Delta (x) =\Delta$ we make here is not the same as the BCS mean field theory in which one essentially assumes that $\Psi(x) = \Delta$, a real number whose equilibrium value (for a uniform superconductor) is determined selfconsistently, neglecting phase fluctuations totally. (In our language, this is done by the extremization (actually minimization) of ${\cal F}_0(\Delta)$, namely by requiring that $(\partial {\cal F}_{0}/\partial \Delta)=0$). Our approach is also valid at all temperatures below $T_{c}$ and thus goes beyond the Ginzburg Landau theory which proposes a phenomenological functional $F_{\rm GL}(\Psi)$ valid for small $\Psi$ ( i.e. near $T_{c}$)  which varies smoothly. (As is well known, this was microscopically justified by Gor'kov \cite{Gorkov} who identified $\Psi (x)$ with the (Cooper) pair function $\Delta(x)$) .


\subsection{Phase Dependent Free Energy}

In this paper, we restrict ${\cal F}_\phi$ to harmonic or Gaussian approximation to ${\rm \bf v}_s^2(x)$ and need to evaluate only time-independent $\phi ({\rm \bf r}) $ for purposes of evaluating the superfluid density or stiffness. Equation (\ref{F_phi_1}) thus yields
 \begin{eqnarray}
&& {\cal F}_\phi = \frac{m}{8}\int d{\rm \bf r}\, {\rm \bf v}_s^2({\rm \bf r}) \, {\rm tr} \,[\sigma_3 \langle {\cal G}(x,x')\rangle_{\rm dis}]_{{\rm \bf r}'={\rm \bf r}, \bar{\tau}' = \bar{\tau} + 0^+} \nonumber \\
 & +& \frac{1}{8}  \int d{\rm \bf r}\, {\rm \bf v}_s^2({\rm \bf r}) \, \left[ \int_0^\beta d\bar{\tau}'\int d{\rm \bf r}'\,{\rm tr} \langle {\rm \bf p}_\alpha {\cal G}(x,x') {\rm \bf p}_\alpha {\cal G}(x',x) \rangle_{\rm dis} \right] 
 \end{eqnarray}
 with transverse superfluid velocity, i.e., $\bm{\nabla}\bm{\cdot} {\rm \bf v}_s =0$. Here
 ${\rm tr}$ represents trace in the Nambu basis and $\langle \cdots \rangle_{\rm dis}$ represents configuration average of the quantities for different disorder realizations. 
 The first term in the above equation is the diamagnetic contribution (Fig~1a) which is the London term, the sole nonzero contribution at $T=0$. The second or paramagnetic term describes the momentum-momentum correlation due to two particles of the same initial momentum and frequency moving in the uniform pair potential and the zero range random, `white noise' potential. Because the random potential is of zero range, as is well known, vertex  corrections (a possible process is shown in Fig.~1c), vanish and one can write the configuration averaged two particle Green's function as the product of two configuration averaged one particle Green's functions (Fig.~1b).

Considering space-time independent $\Delta$, time independent $\phi ({\rm r})$, static 
 non-magnetic disorder in Born approximation, and harmonic or Gaussian approximation to ${\rm \bf v}_s^2({\rm \bf r})$, we therefore find free energy functional
 \begin{equation}
 {\cal F}(\Delta, {\rm \bf v}_s) = {\cal F}_0(\Delta)+ \frac{1}{2}\rho_s (\Delta) \int d{\rm \bf r}\, {\rm \bf v}_s^2({\rm \bf r}) 
 \label{Free_energy_final}
 \end{equation} 
 with
 \begin{equation}
 \rho_s (\Delta) = \frac{m}{4} \left( n + \frac{\hbar^2}{3m} \frac{1}{\beta}\sum_{\omega_n}\int \frac{d{\rm \bf k}}{(2\pi)^3}\, {\rm \bf k}^2\,{\rm Tr }\left[  {\cal G}({\rm \bf k},\omega_n) {\cal G}({\rm \bf k},\omega_n) \right]\right)
 \label{Stiffness_1}
\end{equation}
 where ${\rm \bf p} = \hbar {\rm \bf k}$, the angular average of ${\rm \bf k}_\alpha {\rm \bf k}_\alpha = {\rm \bf k}^2/3$, and $n$ is the electron density. In Eq. (\ref{Free_energy_final}),
 the minimum of the first term gives the self consistent mean field BCS value of the gap in the presence of disorder. The superfluid stiffness $\rho_s$ is like the mass-density of the superfluid; $(1/2)\rho_s {\rm \bf v}_s^2$ is its kinetic energy density.
 As the superfluid density is proportional to superfluid stiffness at zero frequency and momentum as determined here, this approach gives us a direct route for obtaining the superfluid density in terms of the properties of the BCS superconducting state.

The above formulation in Ref.\onlinecite{TVR}, the first to describe the free energy of the 
 electron system microscopically in terms of the pair phase degree of freedom, was used along with the exact eigenstates method of de Gennes \cite{deGennes} . It describes $\rho_s$, the coefficient of the second order terms above, in terms of measured conductivity of the system in the presence of the same disorder $V({\rm \bf r})$ but in the absence of pair interaction. The focus there was to investigate the effect of Anderson localization of electronic states on $\rho_s$.


\section{Superfluid Density}

Equating superfluid stiffness (\ref{Stiffness_1}) with superfluid density $n_s(T)$ as $\rho_s = (m/4)n_s(T)$, we find an expression for the superfluid density as
\begin{equation}
n_s(T) = n + \frac{\hbar^2}{3m} \frac{1}{\beta}\sum_{\omega_n}\int \frac{d{\rm \bf k}}{(2\pi)^3}\, {\rm \bf k}^2\,{\rm Tr }\left[  {\cal G}({\rm \bf k},\omega_n) {\cal G}({\rm \bf k},\omega_n) \right]
\label{Sdensity_1}
\end{equation}
By explicit evaluation of $n_s(T)$ in Eq.(\ref{Sdensity_1}) with the use of Eqs.(\ref{Green}) and (\ref{frequency_relation}), we find 
\begin{equation}
n_s(T) = n \left[ 1+ \frac{1}{\beta}\sum_{\omega_n} \int d\xi_{\rm \bf k} \frac{\xi_{\rm \bf k}^2 + \tilde{\Delta}^2 - \tilde{\omega_n}^2}{(\xi_{\rm \bf k}^2 + \tilde{\Delta}^2 + \tilde{\omega_n}^2)^2} \right]
\label{stiff_temp}
\end{equation}
whose zero temperature value can be expressed as
\begin{equation}
n_s(T=0) = n\left( 1 + \int \frac{d\omega}{2\pi}\int d\xi_{\rm \bf k} \left[ \frac{1}{(\xi_{\rm \bf k}^2 + \tilde{\Delta}_0^2 + \tilde{\omega}^2)}  
 - \frac{2 \tilde{\omega}^2}{(\xi_{\rm \bf k}^2 + \tilde{\Delta}_0^2 + \tilde{\omega}^2)^2} \right]\right). 
\end{equation}
Performing integration by parts for the first term in the above integral, the formal divergence factor can be removed \cite{A-G} to obtain
\begin{equation}
n_s(T=0) = n\left[ 1 +  \int \frac{d\omega}{2\pi}\int d\xi_{\rm \bf k}  \frac{2 \tilde{\omega} (\omega-\tilde{\omega})}{(\xi_{\rm \bf k}^2 + \tilde{\Delta}_0^2 + \tilde{\omega}^2)^2} \right] \, 
\end{equation}
which further simplifies to
\begin{equation}
n_s(T=0) = n\left[ 1-\frac{1}{4\tau} \int d\omega \frac{\omega^2}{(\Delta_0^2+\omega^2)\left( \sqrt{\Delta_0^2+\omega^2} +1/2\tau\right)^2 }\right]
\label{Sdensity_2}
\end{equation}
by performing the integration over $\xi_{\rm \bf k}$ and using the relations in Eq.(\ref{frequency_relation}). The integration in Eq.(\ref{Sdensity_2}) yields
\begin{equation}
n_s(T=0) = \left\{ \begin{array}{l} n (\pi \Delta_0 \tau) \left[ 1- (8/\pi)\frac{\Delta_0\tau}{\sqrt{1-(2\Delta_0 \tau)^2}} \tanh^{-1}\left( \sqrt{\frac{1-2\Delta_0\tau}{1+2\Delta_0\tau}} \right) \right], \,\,\, {\rm for }\,\, 2\Delta_0\tau <1 \\
n (\pi \Delta_0 \tau) \left[ 1- (8/\pi)\frac{\Delta_0\tau}{\sqrt{(2\Delta_0 \tau)^2-1}}   \tan^{-1}\left( \sqrt{\frac{2\Delta_0\tau-1}{2\Delta_0\tau+1}} \right)     \right], \,\,\, {\rm for }\,\, 2\Delta_0\tau >1
\end{array}    \right. \, ,
\label{Sdensity_3}
\end{equation}
 in agreement \cite{Kogan} with the result obtained using Eilenberger quasiclassical limit \cite{Kogan2}   of the BCS theory.

In the pure case $(\Delta_0\tau \gg1)$, we recover the London limit for the superfluid density, i.e., $n_s(T=0) = n$ and in the extreme impure limit $(\Delta_0\tau \ll 1)$, superfluid density is linear in $\tau$, i.e., $n_s(T=0) = n\pi \Delta_0 \tau$, as is well known \cite{deGennes} (this is the `dirty superconductor' limit). 
 Figure 2 shows the variation of $n_s(T=0)$ with $\Delta_0\tau$ obtained using Eq.(\ref{Sdensity_3}). We see that its rate of increase with $\Delta_0\tau$ gradually slows down from linear at small $\Delta_0\tau$ to exponentially small at large $\Delta_0\tau$, reaching the asymptotic London limit. 

Since the experimental data of $n_s(T=0)$ are usually available as a function of conductivity, $\sigma$, which is associated with the dimensionless parameter involving Fermi energy, namely $\epsilon_{_F} \tau$, rather than $\Delta_0 \tau$, we show its variation in Fig.~3(a) with $k_{_F}\ell = (2\epsilon_{_F}/\Delta_0) \Delta_0 \tau$ for two specific values of $\Delta_0/\epsilon_{_F}$ in the ball park of the experimental regime, where $k_{_F}$ and $\ell$ are Fermi wave number and mean free path respectively for electrons. We note that the value of $n_s$ is approximately within $1$--$10$\% of its London limit in the usual experimental range of $k_{_F}\ell$. The linear behavior at small $k_{_F}\ell$ satisfactorily agrees with the experimental data \cite{Pratap1} (Fig. 3(b)) which has been shown as the variation with $\sigma$ which is proportional to $k_{_F}\ell$.

\subsection{Superfluid density at nonzero temperature}

The superfluid density at a nonzero temperature has a generalized form derivable using Eq.(\ref{Sdensity_2}) as
\begin{equation}
n_s(T) = n\left[ 1 +\frac{\pi}{2\tau} \int \frac{dz}{2\pi i} \frac{z^2}{(\Delta^2-z^2)\left(\sqrt{\Delta^2-z^2}+ 1/2\tau\right)^2}\, \frac{1}{e^{\beta z}+1}  \right]
\label{Sdens_T1}
\end{equation} 
where the integration is in the complex plane and $1/(\exp[\beta z]+1)$ is the Fermi function and $\Delta$ is the temperature dependent BCS gap $\Delta(T)$. The associated integrand has poles at $z = \pm \Delta$ and branch cuts for $ \Delta < z$ and $ z < -\Delta$ along the appropriate contour. Calculating the residues at the poles and subtracting the contributions along above and below the branch cuts we find in terms of a real integral,
\begin{eqnarray}
n_s(T) &=& n\left[ 1 + \pi\Delta \tau\tanh\left( \frac{\Delta}{2T}\right) \right. \nonumber \\
&& \left. - \frac{\pi}{\tau^2} \int_{\Delta}^\infty  \frac{d\epsilon}{2\pi} \frac{\epsilon^2}{\sqrt{\epsilon^2 -\Delta^2} \left( \epsilon^2 -\Delta^2 
	+ 1/(2\tau)^2\right)^2 } \tanh\left( \frac{\epsilon}{2T}\right)  \right] \, .
\label{Sdens_T2}
\end{eqnarray}
As expected, the zero-temperature limit $(T\to 0)$ of Eq.(\ref{Sdens_T2}) reproduces $n_s$ as shown in Fig.~2. Figure 4 shows the temperature dependence of $n_s$ obtained using Eq.~(\ref{Sdens_T2}) for different values of ${\rm k}_{_F}\ell$.  

It is convenient to expand Eq.~(\ref{stiff_temp}) in the power series of $\Delta^2(T) $ for the purpose of obtaining $n_s(T)$ near $T_c$ as $\Delta^2(T) \sim (8\pi^2/7\zeta(3)) (T_c -T)T_c$ is very small, where $\zeta(3)$ is a Riemann zeta function. We thus find
\begin{equation}
n_s(T\sim T_c) \approx n\Delta^2\frac{\pi}{\beta} \sum_{\omega_n}  \frac{1}{\omega_n^2 (\vert \omega_n \vert+ 1/(2\tau))}
\end{equation}
By employing the standard algebra for the series sum with Matsubara frequency, we find
\begin{eqnarray}
n_s(T\sim T_c) &=& n\left( \frac{\Delta}{2T_c} \right)^2 \left[ (1/\pi) (4T_c\tau) \Psi'\left( \frac{1}{2}\right) \right. \nonumber \\
&& \left. - (4T_c\tau)^2 \left\{ \Psi \left( \frac{1}{2} +\frac{1}{4\pi T_c\tau}\right) - \Psi\left( \frac{1}{2}\right)  \right\}  \right]
\end{eqnarray}
where $\Psi(x)$ is the digamma function and $\Psi'(x)$ is its derivative. It is easy to check that for a clean superconductor ($T_c\tau \to \infty$), $n_s(T\sim T_c) = 2n(T_c-T)/T_c$ as known. The superfluid density is proportional to $\Delta^2(T)$ near $T_c$ and the proportionality constant decreases with the increase of disorder. 


\section{Discussion}

We have described above the first general microscopic approach for determining the free 
 energy functional ${\cal F}(\Delta(x),\phi(x))$ of a superconductor which involves explicitly the amplitude  $\Delta (x)$ of the order parameter, \textit{as well as} its phase $\phi(x)$. This is in content quite different, though similar in broad spirit as the phenomenological Ginzburg-Landau functional \cite{GL} and its microscopic justification by Gor'kov \cite{Gorkov} through a power series expansion of $ \Psi({\rm \bf r}) = \Delta ({\rm \bf r})e^{i\phi({\rm \bf r})}$. The latter, being an expansion for small $ \Psi({\rm \bf r})$, is valid only in the neighborhood of $T_c$. The functional here consists of all orders in $\Delta (x)$ and up to the desired order in $\phi (x)$ (of the gradient in $\phi(x)$ to be precise) and is thus valid for all temperatures. Because it involves the phase explicitly, this approach will be useful for superconductors in all kinds of situations in which the phase plays an important role. Likely possibilities are disordered superconductors whose coherence lengths are necessarily short, because of which fluctuation effects are significant. Another general context is coulomb interactions which, as is well known \cite{TVR}, couple to time dependent phase fluctuations. One can think of inhomogeneous situations such as certain kinds of boundaries and impurities where the phase changes with distance, and quantum dot like geometries with locally enhanced coulomb interactions.  Generalized varieties of Josephson junctions in which phase is the critical degree of freedom, are also a possible domain of application. In this paper, we have applied it to perhaps the simplest case, namely the superfluid stiffness of a BCS s-wave superconductor in presence of non-magnetic disorder, a dc property, in the Gaussian or harmonic approximation (where the fluctuations with different wavevector ${\rm \bf q}$ and frequency $\omega$ are independent).

 We have described above a general approach to the superfluid density $n_{s}$ of superconductors 
 as a function of disorder. The approach is based on describing the energy of the superconductor as a function of the superfluid velocity. The stiffness is the (half of the) coefficient of the square of the superfluid velocity, a gauge invariant quantity. The superfluid stiffness (or equivalently, density $n_{s}$) is directly measured through the penetration depth. We have calculated $n_{s}$ both at $T=0$ and $T\neq 0$. It has the expected pure (London) limit at $T=0$ and vanishes appropriately as $T$ approaches $T_{c}$. We have exhibited closed form expressions for $n_{s}(T)$ for all $T$ as a function for a wide range of $ \tau$, from the very clean limit $(\Delta_0\tau>>1)$ to the normal disordered metal regime $\epsilon_{F}\tau >1$, but $\Delta_0\tau>1$  or $<1$ (these results are available for the first time). We have compared our results with experiment.

The approach developed here has the immediate  potential for exploring many open questions concretely 
 by using 
 Eqs. (\ref{F_phi_1}) and (\ref{H_phi_1}). These are basically questions where the effect of the phase of the superconductive order is important. Experimentally, this seems to the case when there is significant static disorder and (or) the effective coulomb interactions are strong.  An evaluation of the leading anharmonic (quartic) terms in the phase functional, and even a quasi harmonic approximation to it, leads to a requirement for a selfconsistent calculation of $\rho_s$ at all temperatures and disorder, and therefore to a disorder-dependent temperature $T_{\phi}$ above which $\rho_s$ or phase stiffness vanishes. This temperature is lower than the mean field, BCS $T_{c}$ which is the temperature at which the amplitude $\Delta$ of the order parameter becomes nonzero, namely at which zero energy Cooper pairs form. One thus has a temperature regime with nonzero $\Delta$ but vanishing phase stiffness (a non-superconducting pseudogap (?) regime) which becomes larger with increasing disorder, for example. 
  One can also calculate the phase propagator in this regime and determine its effect on the nature of the single particle states, in particular on the single particle density of states, and thus address the question of the pseudogap. The renormalization of superfluid stiffness by fluctuations involves not only space dependent phase fluctuations via the superfluid velocity, but also time dependent phase fluctuations which become stronger with increasing strength of the effective coulomb interaction. Because of this, the quantum (time dependent) phase fluctuation induced reduction of the phase stiffness becomes larger with increasing Coulomb interactions, and the renormalized stiffness may go to zero! \cite{footnote_new} This may explain the phenomenon  of `failed superconductivity' reviewed in Ref.\onlinecite{Kivelson}. The observed vortex lattice melting \cite{Ong,Pratap3} at low temperatures is most likely related to large quantum fluctuations of the phase.

\begin{figure}
	\includegraphics[scale=0.6]{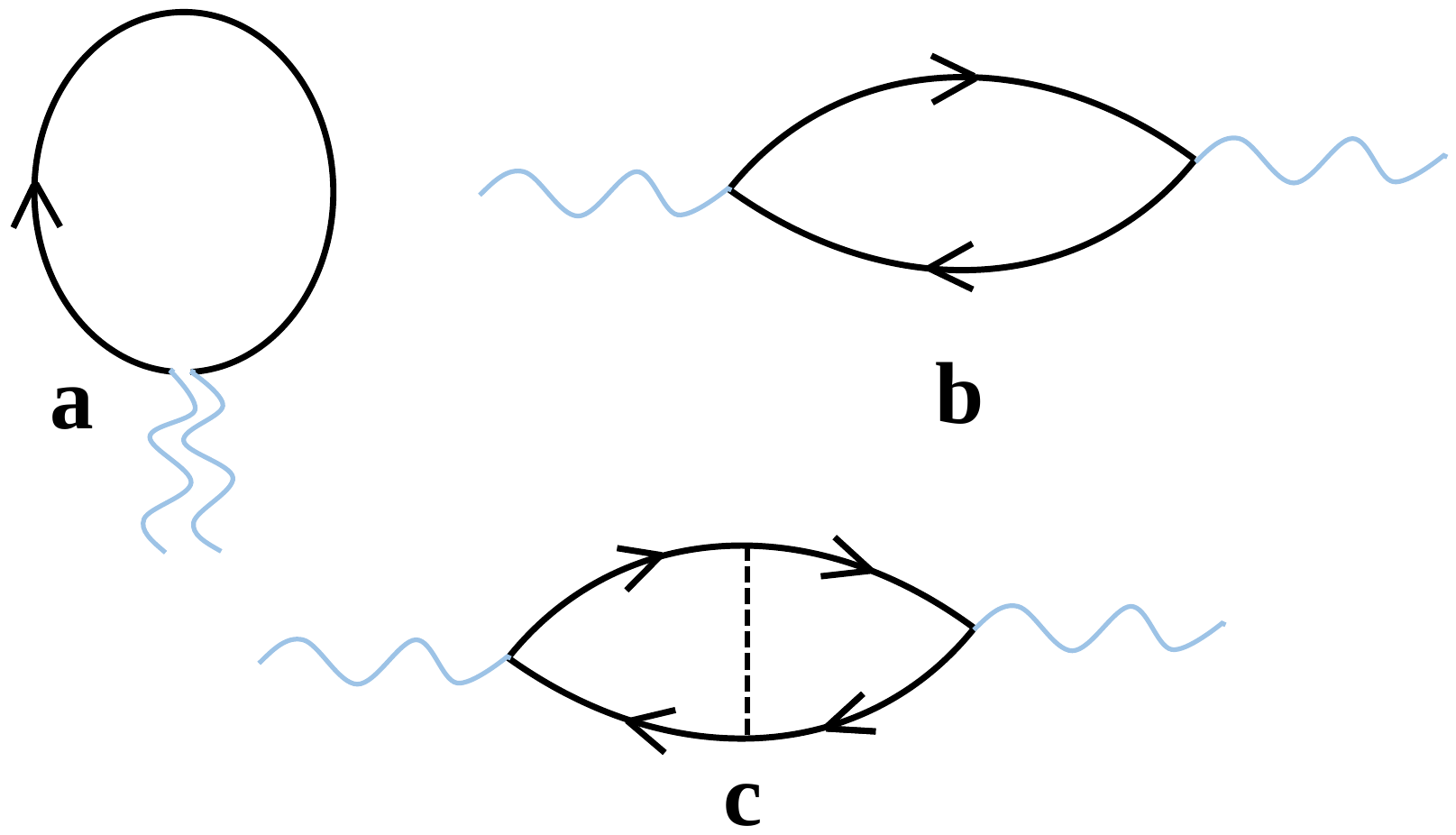}
	\caption{(color online) Feynman diagrams for the contributions to superfluid density. While wavy lines represent the superfluid velocity, the solid lines represent the Fermionic Nambu Green's functions. (a) The diamagnetic contribution. (b) The paramagnetic contribution without vertex correction. (c) The paramagnetic contribution from vertex correction due to one scattering (represented by dashed line) of Nambu quasiparticles and holes from the same scattering center; this however vanishes for the short-range disordered potential considered here. The vertices with superfluid velocity and fermions in (b) and (c) are momentum vertices.}
\end{figure}

\begin{figure}
	\includegraphics[scale=0.6]{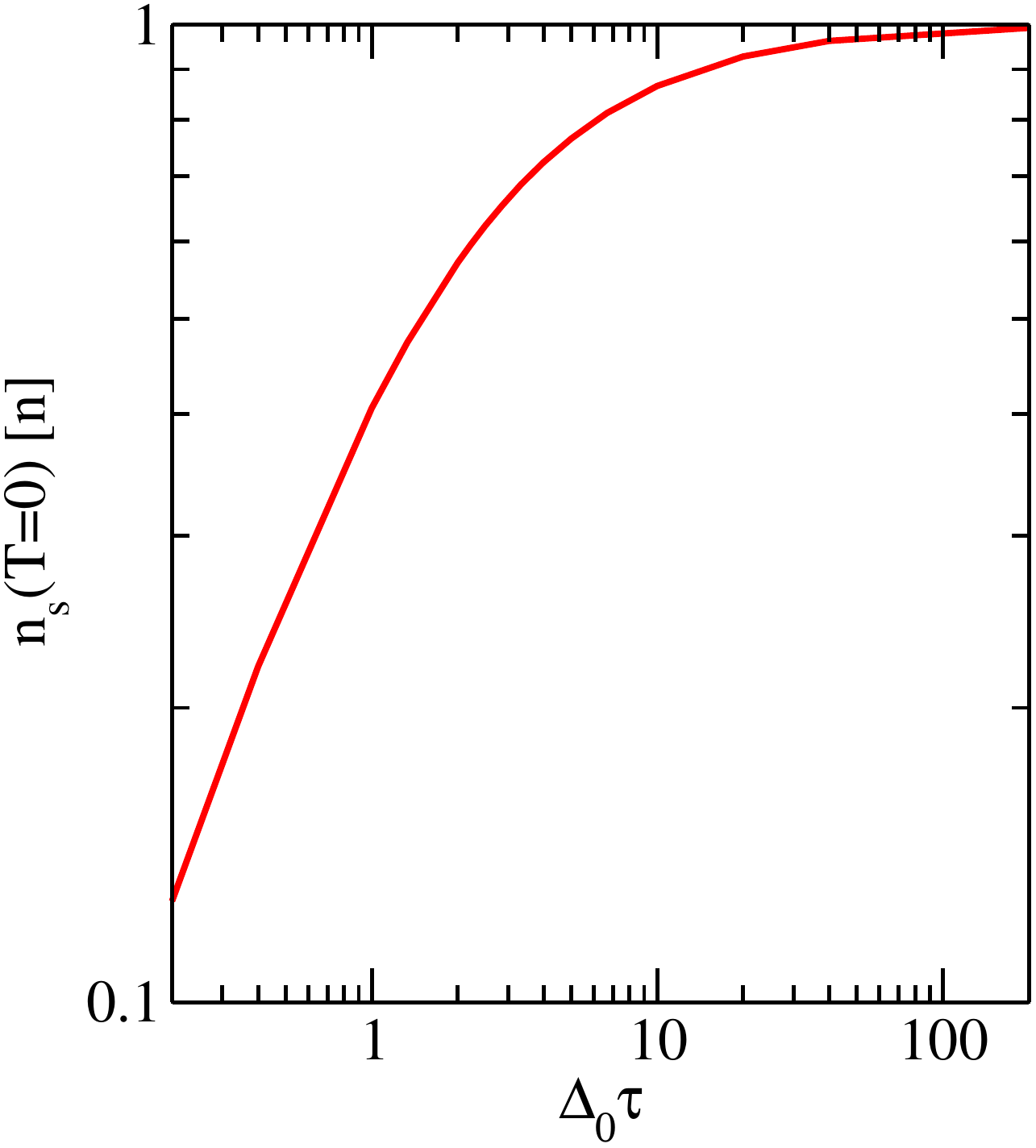}
	\caption{(color online) The superfluid density at zero temperature versus $\Delta_0 \tau$ obtained using Eq.(\ref{Sdensity_3}).}
\end{figure}

\begin{figure}
	\includegraphics[scale=0.6]{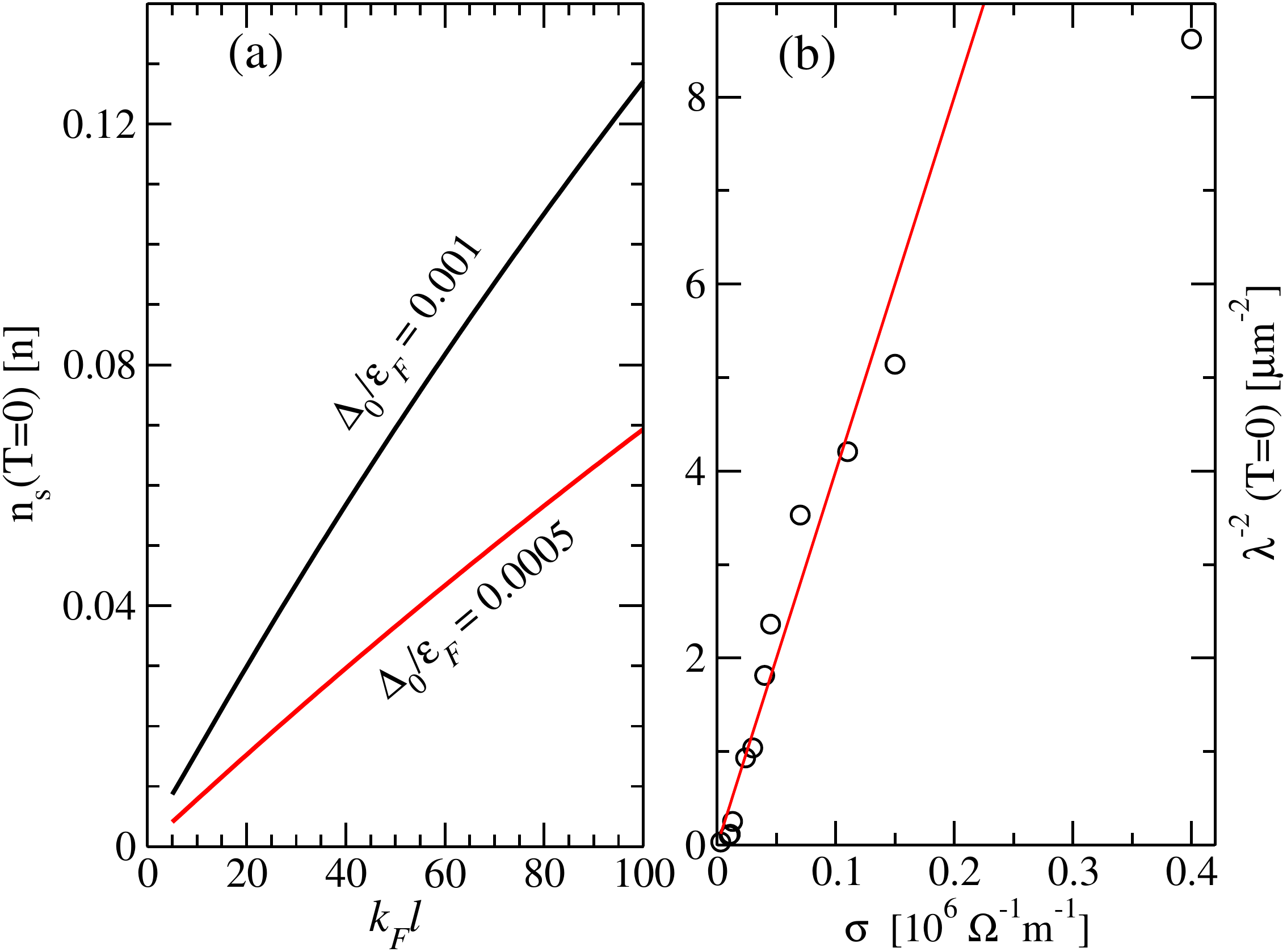}
	\caption{(color online) (a)The superfluid density at zero temperature versus $k_{_F}\ell$ for $\Delta_0/\epsilon_{_F} =0.001$ and 0.0005. (b) Experimental data of inverse square of penetration depth $\lambda^{-2}$ and the corresponding conductivity $\sigma$ extracted from Ref.\onlinecite{Pratap1} where the measurements were performed in epitaxial NbN films with thickness much greater than dirty-limit coherence length. The solid line is a guide to the eye for roughly linear dependence at large disorder. }
\end{figure}

\begin{figure}
	\includegraphics[scale=0.6]{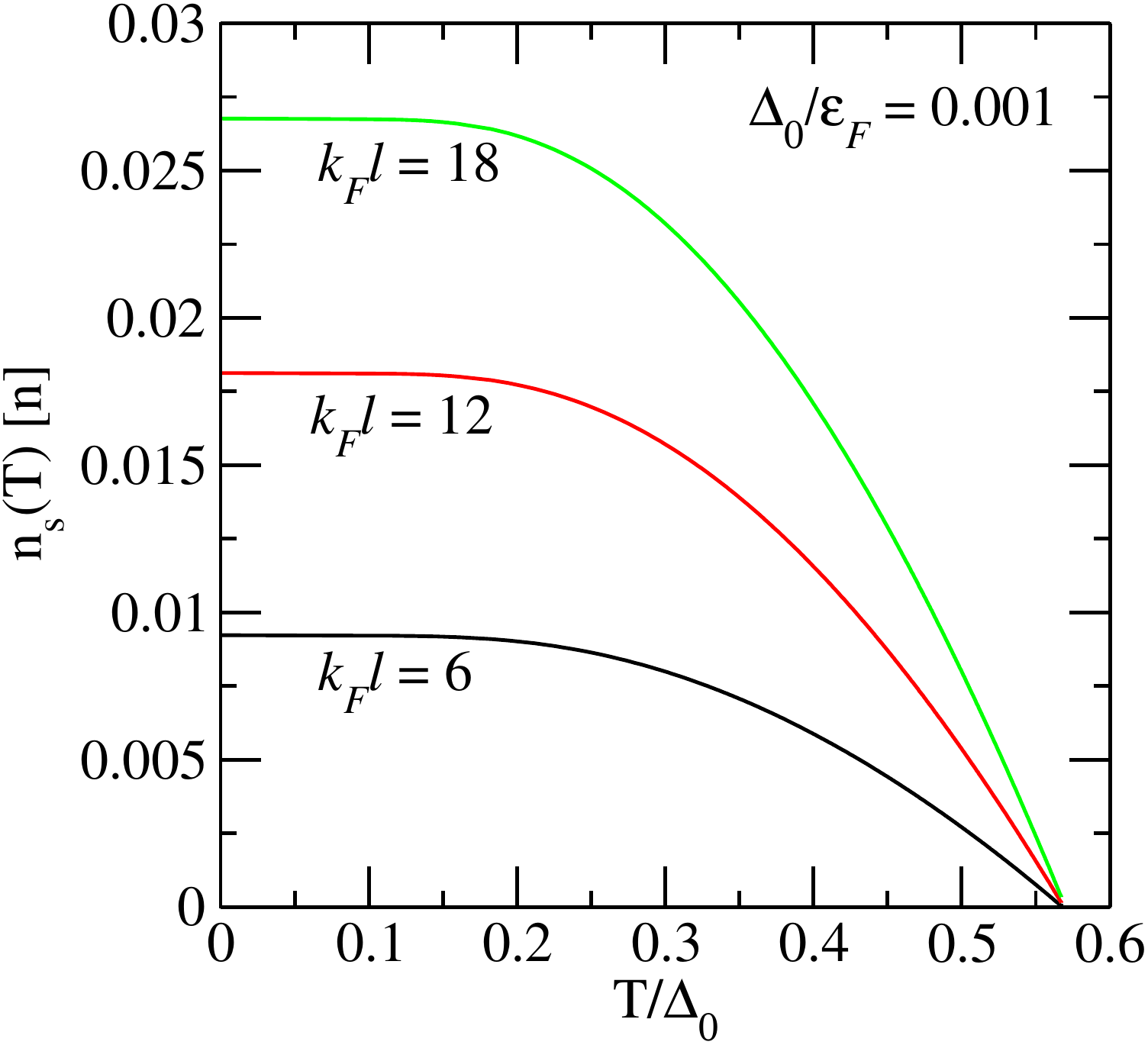}
	\caption{(color online) Superfluid density versus temperature at different values of ${\rm k}_{_F}\ell$. The critical  temperature for vanishing superfluid density coincides with the BCS $T_c$. }
\end{figure}



\begin{thebibliography}{99}
\bibitem{footnote_new}The very fact that $T_{\phi}$ depends on disorder while the BCS $T_{c}$ does not, indicates that there must be such a regime. A crude analogy is with magnetic ordering of spin systems. The interacting spins order at a temperature which is much lower than the temperature at which the spins form. Roughly, the former is analogous to the nonzero phase stiffness temperature of a superconductor and the latter to the temperature at which Cooper pairs form. In the BCS theory, the latter is argued to be indistinguishably higher than the former. The former is taken to be the $T_{c}$. The difference increases as fluctuation effects increase; this happens as the superconducting coherence length $\xi$ decreases, which it does with disorder. 
                   Coulomb interactions between charge fluctuations are proportional to time dependent fluctuations of the Cooper pair phase. Therefore as the coulomb interaction increases, temporal fluctuations of the phase become stronger, and this will reduce the phase stiffness. One can speculate that it could be driven to zero. The question is whether such a phase can be identified with the failed superconductor \cite{Kivelson} which has been observed in many systems and has a well defined set of properties.
\end{thebibliography}
\end{document}